# Ridge Resonance: A new Resonance Phenomenon for Silicon Photonics Harnessing Bound States in the Continuum

*Thach G. Nguyen[1,*], Guanghui Ren[1], Steffen Schoenhardt[1], Markus Knoerzer[1], Andreas Boes[1], and Arnan Mitchell[1,*]*

*Corresponding Author: E-mail: thach.nguyen@rmit.edu.au, arnan.mitchell@rmit.edu.au

[1]School of Engineering, RMIT University, Melbourne, VIC 3001, Australia

**Abstract:** We present a new resonant behavior based on bound states in the continuum in a guided wave silicon platform. The continuum has the form of a collimated beam of light which is confined vertically in a TE mode of a silicon slab. The bound state is a discrete TM mode of a ridge on the silicon slab. The coupling between the slab and ridge modes results in a single sharp resonance at the wavelength where they phase match. We experimentally demonstrate this phenomenon on a silicon photonic chip using foundry compatible parameters and interface it on-chip to standard single mode silicon nanowire waveguides. The fabricated chip exhibits a single sharp resonance near 1550 nm with a line width of a few nanometer, an extinction ratio of 25 dB and a thermal stability of 19.5 pm/°C. We believe that this is the first demonstration of bound states in the continuum resonance realized using guided wave components.

# 1. Introduction

Silicon photonics has history in research spanning the last 30 years, but we are now seeing its emergence as a mature platform with standard building blocks that can be mass manufactured on wafer scale using CMOS foundry technology. One of the most important building blocks of silicon photonics are optical resonators. These are crucial for wavelength filters and routers,[1] resonant modulators [2] and laser cavities [3] among many other applications. Microresonators are of particular industrial significance for integrated data communications systems, providing compact, energy efficient and high-performance modulators and narrow-band filters for wavelength division multiplexing (WDM) functions on chip. Sensors are also emerging as a major industrial application of microresonators.

Recently, the photonics research community has begun to explore alternate resonance phenomena based on bound states in the continuum (BIC).[4] First proposed in 1929 by Neumann and Wigner, these unconventional states can be applied to a large variety of systems and have since been used to describe effects in electromagnetics, electronics and even acoustics.[5] Many recent demonstrations have shown bound states in the continuum for dielectric metasurfaces [6-8]. Here relatively weak resonances with a quality factor on the order of 2000 are deemed of interest with applications demonstrated in biosensing [9] and beam control among others. In guided wave photonics, it has been predicted that BIC can be formed by symmetry-protected bound states [10], radiation cancellation from two or more coupled resonators [11] or radiation cancellation from a single resonance such as in photonic crystal slabs [12], a low-index waveguide on a high-index membrane [13]. The implementation of such structures in photonic integrated circuits and especially in the most common platform of silicon photonics is therefore very appealing as they could be exploited to achieve robust and high extinction integrated wavelength filters with some of the attractive properties of thin-film filters.

In this article, we experimentally demonstrate a bound state in the continuum resonance in an integrated silicon photonic waveguide platform. This resonance is achieved by coupling a continuum of un-bound TE slab modes to a single propagating TM mode of a ridge. We thus call this *ridge resonance*. We present the theory underpinning this resonance effect identifying the parameters that determine the resonance characteristics. We then present proof of concept experimental results on the fabricated device, demonstrating a single sharp resonance – exhibiting remarkably deep resonance and low temperature dependence. We believe that this is the first experimental demonstration of waveguide based bound state in the continuum resonance of any type in silicon photonics. However, our demonstration is made more significant as it is achieved using only foundry compliant parameters and interfaced to standard silicon photonic nanowire waveguides. While the demonstrated resonance is relatively broad, it would be well suited to coarse wavelength multiplexing and sensing applications – similar to the metasurface counterparts expanding the toolbox of filtering approaches for silicon photonics.

**2. Lateral leakage of TM mode in a ridge: leakage of the bound state into the continuum**

To introduce our silicon photonic bound state in a continuum, we will first explain how bound states can couple to the continuum on the silicon platform. Consider a silicon ridge as illustrated in **Figure 1(a)**. This ridge can support guided modes in both TE and TM polarizations. The TE mode of such a ridge propagates with low loss and is commonly used in silicon photonics for optical transport.[14,15] Our team discovered that the TM mode can radiate strongly and equally to the TE slab mode on either side (as illustrated in **Figure 1(b)**), an effect called lateral leakage.[16-20] The generated TE slab mode propagates at a particular angle to the waveguide axis given by

$$\theta = \cos^{-1}\left(\frac{n_{tm}}{n_{te}}\right) \quad (1)$$

where $n_{tm}$ and $n_{te}$ are the effective index of the TM guided mode and TE slab mode in the claddings, respectively.

The strength of radiation at each wall is determined by the geometry of that wall.[21] The total radiation is the coherent sum of radiation from each wall and hence the wall separation can cause resonant enhancement or suppression of the leakage. Therefore, by carefully designing the waveguide width, it is possible to achieve very low propagation loss for TM mode.[16, 22] The width at which complete cancellation occurs has been termed the 'magic' width.[22] This 'magic' width can be readily found using a closed-form expression based on the effective index of the TE slab mode in the ridge and the TM guided mode.[16,22] A similar effect has also been discovered in disk and ring resonators in which the leakage loss depends not only on the ridge width (for the case of rings) but also on the disk or ring radius.[19,20] These effects of low loss propagation due to leakage cancellation can also be considered as guiding light through optical bound states in the continuum (BIC).[8] The TE waves radiated from both sides of the ridges are loss channels. The phase difference between them are controlled by the ridge width (and/or radius for the case of rings/disks) to achieve cancellation of loss to the continuum. It has been predicted theoretically that similar lateral leakage due to the coupling between a bound TM mode and a continuum of TE slab mode and leakage cancellation or BIC behavior can also be achieved with a low refractive index waveguide on a high refractive index thin membrane.[13] However, to our knowledge, these latter forms of leakage have yet to be experimentally demonstrated.

**3. TE slab beam incident on a ridge: excitation of the bound state from the continuum**

Our previous works have extensively explored leakage from a bound state into the continuum in silicon thin-ridges [22]. In this paper, we explore the reverse effect. Consider the situation illustrated in **Figure 1(c)** where the TE slab mode is excited as a broad beam, which is incident on the ridge. To a first order approximation, one might expect the beam to experience Fresnel reflection at the rising and falling edges of the ridge due to the change in effective index of the TE mode in the slab and ridge regions. An estimate of the reflection as a function of the angle of the incident TE beam that would be expected considering only Fresnel reflection is depicted in **Figure 1(d)** labelled 'Background' (refer to **Supporting Information, Section 1** for details). However, this approximation does not describe what will actually be observed as it ignores coupling between the TE beam in the continuum and the bound TM mode of the ridge. A full vector simulation (**Supporting Information, Section 2**) which rigorously takes into account the TE/TM coupling reveals the true behavior as illustrated in **Figure 1(d),** labelled 'Rigorous'. It can be seen that rigorous calculation shows a strong resonance with 100% reflection and far away from this resonance, the reflection is approximately equal to the background.

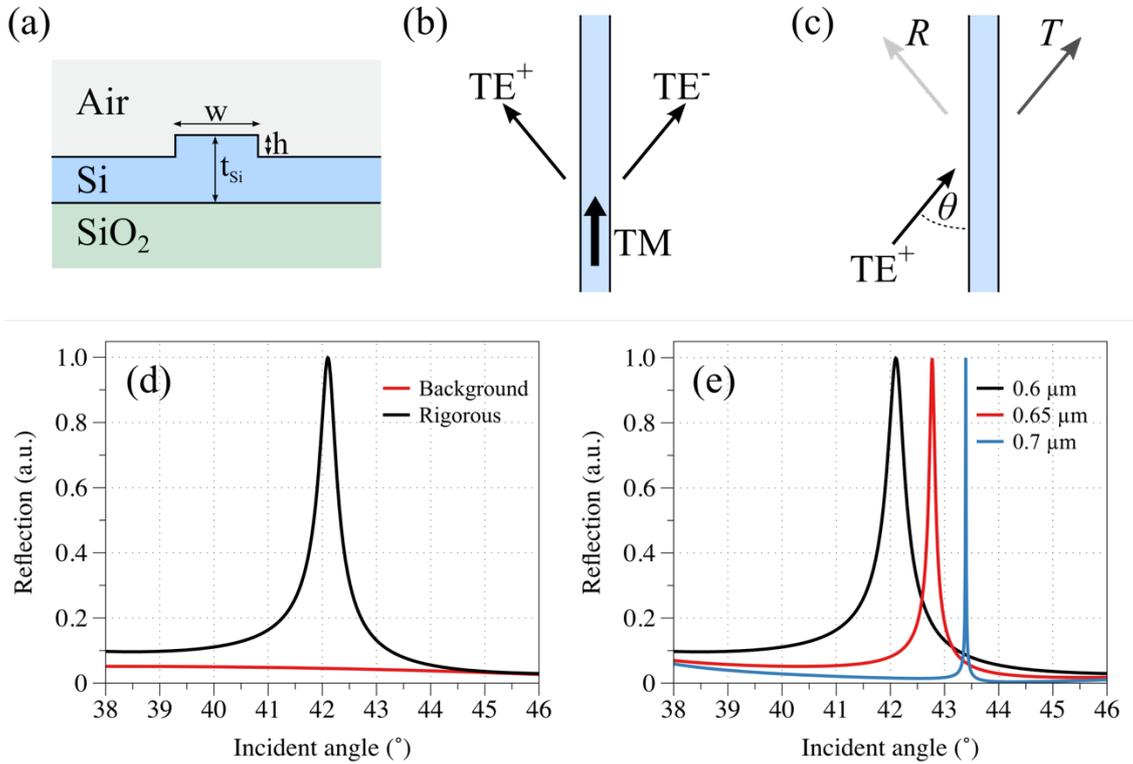

**Figure 1.** (a) Schematic of the ridge cross-section; silicon thickness $t_{Si}$ is 220 nm and the etching depth h is 70 nm. (b) Illustration of a ridge, which leaks TE slab modes on either side when excited with an TM mode. (c) Illustration of the reverse process, where the ridge is illuminated with a TE slab mode from one side, resulting in reflected $R$ and transmitted $T$ TE light. (d) Reflection as a function of the incident angle $\theta$ when the ridge width w = 0.6 μm, showing the Fresnel background and rigorous simulation solutions. (e) reflection as a function of the incident angle for different ridge widths.

**Figure 1(e)** shows the rigorously simulated reflection as a function of the incident angle for different ridge widths. It can be shown that the angles at which 100% reflection occurs are equal to the angles calculated using **Equation (1)**. At these angles, the incident TE slab mode is phase matched to the TM guided mode in the ridge. Therefore, it is possible that the TM guided mode is excited in the ridge at these angles. To reveal that this is indeed the case, we plot the field distributions when a TE optical field is launched toward the ridge with an incident angle that is on resonance and 4° different to the on resonance incident angle. **Figure 2(a)** and **Figure 2(b)** show the interaction of the optical field with the resonator (which is located at the zero position of the x-axis) in the bird view perspective (similar to **Figure 1(b)**). **Figure 2(c)** and **Figure 2(d)** show the optical field interaction in the cross-section view of the ridge resonator. Horizontal electric field component ($|Ex|$) is plotted in **Figure 2(a)** and **Figure 2(b)**, while vertical electric field component ($|Ey|$) is shown in **Figure 2(c)** and **Figure 2(d).** On resonance, it can be seen that the TE beam interacts strongly with the TM mode of the ridge, building up power over the width of the beam and then gradually re-radiating the beam as strong reflection. Off resonance, there is almost no interaction with the TM mode of the ridge and most of the power of the launched TE beam is transmitted.

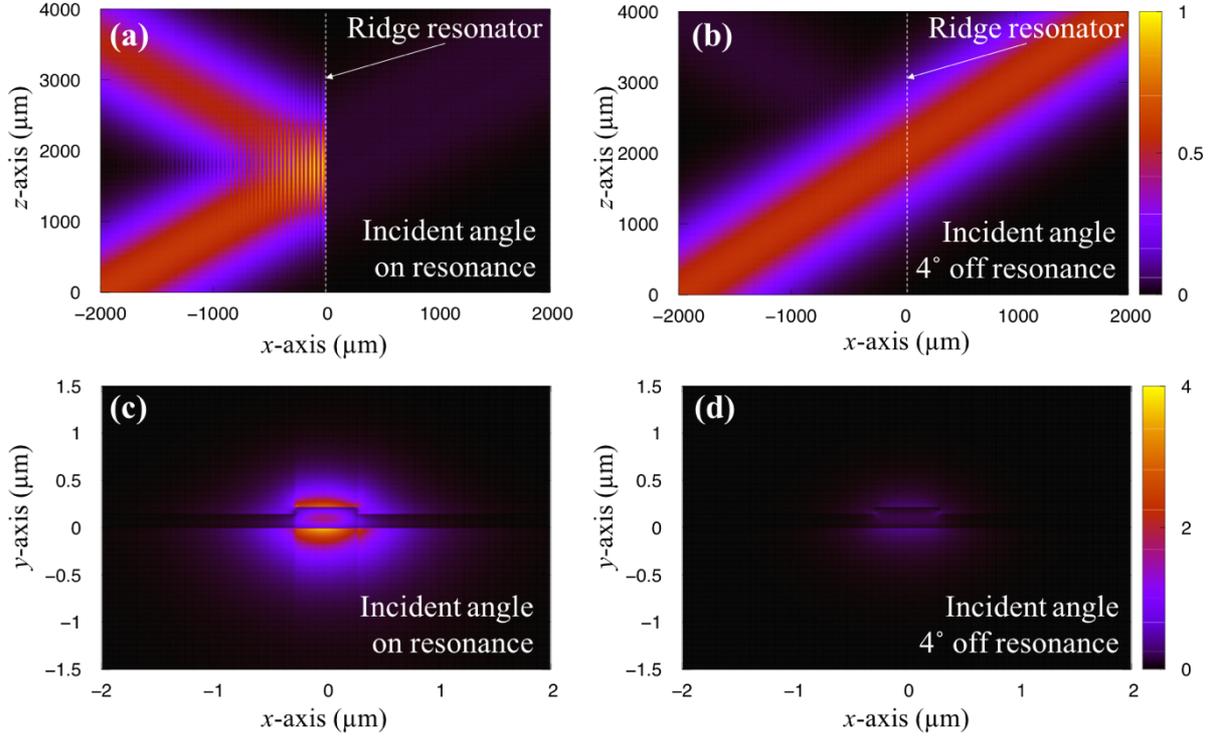

**Figure 2.** (a) and (b) show the simulations of the optical field when the TE slab mode is launched with an incident angle that is on and off resonance in birds view perspective, respectively. The dashed line indicates the ridge resonator position. (c) and (d) present how the optical field interacts with the ridge in the cross-sectional view. The optical field scales are arbitrary units.

It can be seen from **Figure 1(e)** that the angular bandwidth of the reflection depends strongly on the ridge width. When the ridge width approaches the 'magic' width at which the lateral leakage loss of the guided TM mode is zero [16, 18, 22], the angular bandwidth of the reflection also approaches zero. This phenomenon indicates the presence of bound states in the continuum [23] when the ridge at the 'magic' width is excited with a TE beam at an angle resulting in phase matching between the exited TE beam and TM guided mode in the ridge.

In the above analysis, the wavelength is fixed. When the wavelength changes, the incident TE beam in the slab and the guided TM mode in the ridge will no longer be phase matched since the TE slab mode and TM guided mode have different dispersion characteristics. Therefore, it is expected that the reflection and transmission will also be strongly wavelength dependent.

**Figure 3(a)** and **Figure 3(b)** show the transmission and reflection as a function of wavelength calculated by the rigorous simulation method (refer to **Supporting Information, Section 2** for details), respectively when the incident angle of the TE beam is fixed. Also shown in **Figure 3(a)** and **Figure 3(b)** are the reflection and transmission that would be expected considering only Fresnel reflection, labelled "Background". It is evidence that the ridge acts like a resonator to the incident TE beam. At the wavelength where the TE beam is phase matched to the TM mode in the ridge, the reflection is 100%. Far away from that resonant wavelength, the reflection approaches the background Fresnel reflection.

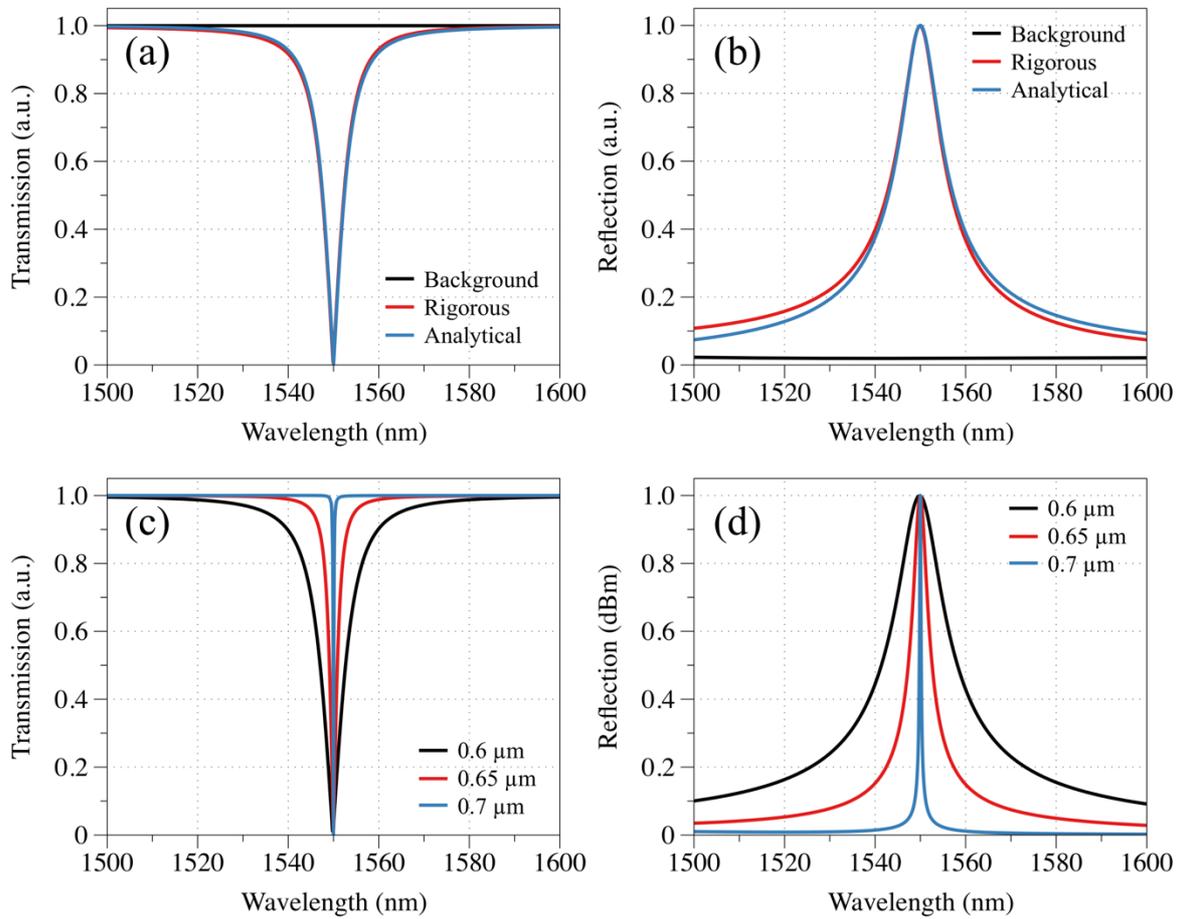

**Figure 3.** (a) Transmission and (b) reflection as a function of wavelength, showing the Fresnel background, rigorous simulation and analytical solution. (c) Transmission and (d) reflection as a function of wavelength, for different ridge widths.

Similar to the case of the angular bandwidth, the spectral bandwidth of the resonance also depends strongly on leakage rate of the guided TM mode. Therefore, it is expected that the ridge width has a strong impact on the resonance width or Q-factor. **Figure 3(c)** and **3(d)** show the transmission and reflection of the ridge resonator as a function wavelength for a ridge resonator width of 0.6, 0.65 and 0.7 μm (and with other parameters defined in Figure 1). One can see that the spectral bandwidth decreases when the ridge width increases from 0.6 to 0.7 μm. Infinite narrow linewidth resonance can be achieved when the leakage rate of the guided TM mode approaches zero, or the ridge is at the 'magic' width.

**4. Coupled mode analysis of the resonance**

It is possible to draw a parallel with the Fano resonance in other BIC photonic systems.[24,25] The full transfer function of an excitation beam on the ridge will be composed of the superposition of a background component (due to Fresnel reflections) and a resonance describing the gradual coupling of the TE beam into the TM bound mode and the reciprocal leakage of this energy back into the TE continuum. Following Wang et al.,[24] it is possible to derive a closed form expression enabling accurate prediction of the resonant spectral response of a beam reflection from a ridge based on propagation behavior of TM mode and the Fresnel coefficients of the ridge walls. The reflection coefficient of a TE beam on a ridge can be expressed as (see **Supporting Information, Section 1** for detailed derivation):

$$R = \frac{j[(n_{te}\cos\theta - n_{tm})r - n'_{tm}t]}{j(n_{te}\cos\theta - n_{tm}) + n'_{tm}} \qquad (2)$$

where $r$ and $t$ are the background reflection and transmission coefficient of the non-resonant process, $n_{tm} + jn'_{tm}$ is the complex effective index of the leakage TM mode. When the background non-resonant reflection is negligible, ie $r = 0$, $t = 1$, the reflection coefficient can be simplified to:

$$R \approx \frac{-jn'_{tm}}{j(n_{te}\cos\theta - n_{tm}) + n'_{tm}} \qquad (3)$$

It can be seen that the spectral characteristic of the resonance is determined by the leakage loss of the TM guided mode as well as the dispersion of the TE slab mode and TM guided mode. 100% reflection occurs at the resonant wavelength or incident angle $\theta$ that satisfies the phase matching condition $n_{te} \cos \theta = n_{tm}$. The quality factor of the resonance increases when the ridge approaches the 'magic' width where the lateral leakage cancellation occurs, for example $n'_{tm} = 0$. When the background reflection is ignored, the quality factor of the resonance is infinite at the 'magic' width, confirming the existence of a BIC.[23] The analytic response as calculated by Eq. (2) is presented in **Figure 3(a)** and **Figure 3(b)** labelled 'Analytical' showing a very close match to the rigorous, full vector simulation result.

In this section, we have provided an analytical analysis showing the dependence of the resonance on the properties of the guided TM mode and TE slab mode including TM leakage loss and the dispersion of TM guided mode in the ridge and TE slab mode in the claddings. These parameters are all determined by the geometry of the ridge including silicon thickness, etch depth and ridge width. How these geometrical parameters impact the properties of the guided TM mode and TE slab were extensively investigated in our previous work.[16,18-20,22]

## 5. Experimental Results

As a proof of concept demonstration, we fabricated a system to launch a Gaussian beam at a ridge. We designed this system using the fabrication constraints of a popular multi-project wafer (MPW) foundry passive technology.[26,27] The layout and the components in the SOI chip are shown in **Figure 4**, where **Figure 4(a)** shows an optical microscope image of the chip. The fabrication of the silicon thin-ridge resonators was carried out on a diced piece of SOI wafer with a 220-nm-thick Si guiding layer and a 3-μm-thick SiO$_2$ buried oxide layer. Firstly, the parabolic reflectors and alignment markers were patterned using a Vistec 5000Plus electron beam lithography (EBL) on a 400-nm-thick ZEP520A EBL resist layer spun on the chip. These structures were fully etched using Oxford PlasmaPro 100 reactive ion etching (RIE) system.

Afterwards, grating couplers, waveguides, apertures and ridge resonators were patterned by a second EBL and RIE process, with an etch depth of only 70 nm.

In the following, a brief overview is given to describe the function of the different structures that were used. The laser light is coupled into a single mode optical waveguide from a standard single mode optical fiber, using focused grating couplers (**Figure 4(c)**). The single mode waveguide then launches a slab waveguide mode by adiabatically expanding the width, before being terminated (**Figure 4(b)**). The slab mode expands and interacts with a parabolic reflector, which collimates the slab mode and reflects it in the direction of the ridge resonator. The collimated slab mode interacts with the ridge resonator (**Figure 4(d)**) under an angle of 52.2˚. The ridge resonator width was 650 nm. The transmitted light is focused into the right single mode waveguide by another parabolic reflector. The light is then coupled out of the waveguide into a single mode fiber by a grating coupler.

In the fabricated device, the ridge resonator was formed by partially etching two trenches in either sides of the ridge. The width of the trenches is 20 μm to ensure that the outer edges of the trenches does not interact with the evanescent TM field of the guided TM mode. Therefore, the resonance component of the overall resonator response due to TE/TM coupling is not affected by the trenches. The trenches only change the background component due to having two extra interfaces between etched and unetched silicon slabs.

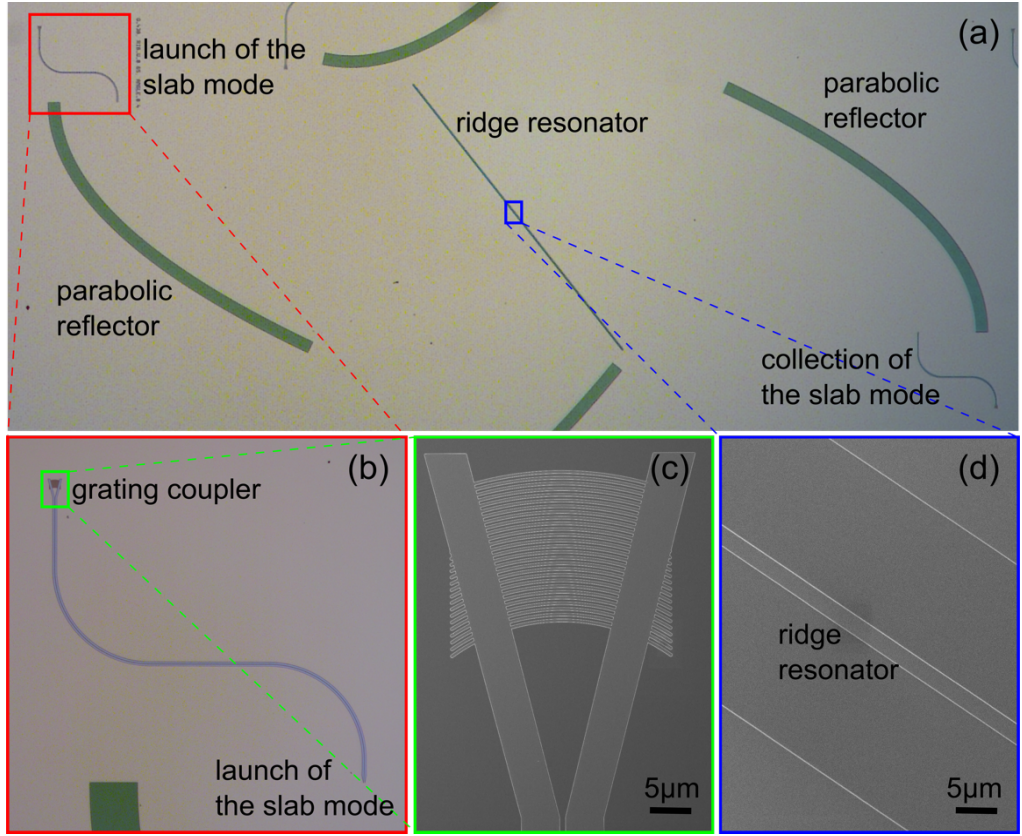

**Figure 4.** (a) Optical microscope image giving an overview of the whole ridge resonator test chip. (b) image of the silicon wire waveguide, which is used to launch a slab mode. (c) SEM image of a focused grating coupler, which is used to couple light into the wire waveguide. (d) SEM image of the ridge resonator.

The ridge resonator structure was characterized by coupling laser light in and out of the chip using the grating couplers and cleaved fibers mounted at an angle of 10°. The positions of the fibers were computer controlled to optimize the transmitted power. A tunable C-band laser (Agilent 81600B) and an in-built optical power detector (Agilent 81635A) were used to measure the transmitted power as a function of the wavelength. The transmission spectrum of this structure is presented in **Figure 5(a)**. The transmission spectrum exhibits a broad envelope, defined by the spectral efficiency of the grating coupler (see Supporting Information, Section 3 for analysis of the system with and without the ridge resonator). Such a spectral response is typical of this type of silicon photonic interface [28]. When the ridge resonator is present, a single, strong resonant dip is evident at wavelength of 1539 nm. The resonant dip is magnified in **Figure 5(b)**. The resonance has an extinction ratio of approximately 25 dB and a full width at half maximum of 2.95 nm, which corresponds to a Q-factor of around 520. This behavior closely matches the predictions of **Figure 3(a)**. It should also be noted that the resonator test structure in **Figure 4(a)** only adds about 6 dB of loss when compared to a straight reference waveguide on the same chip.

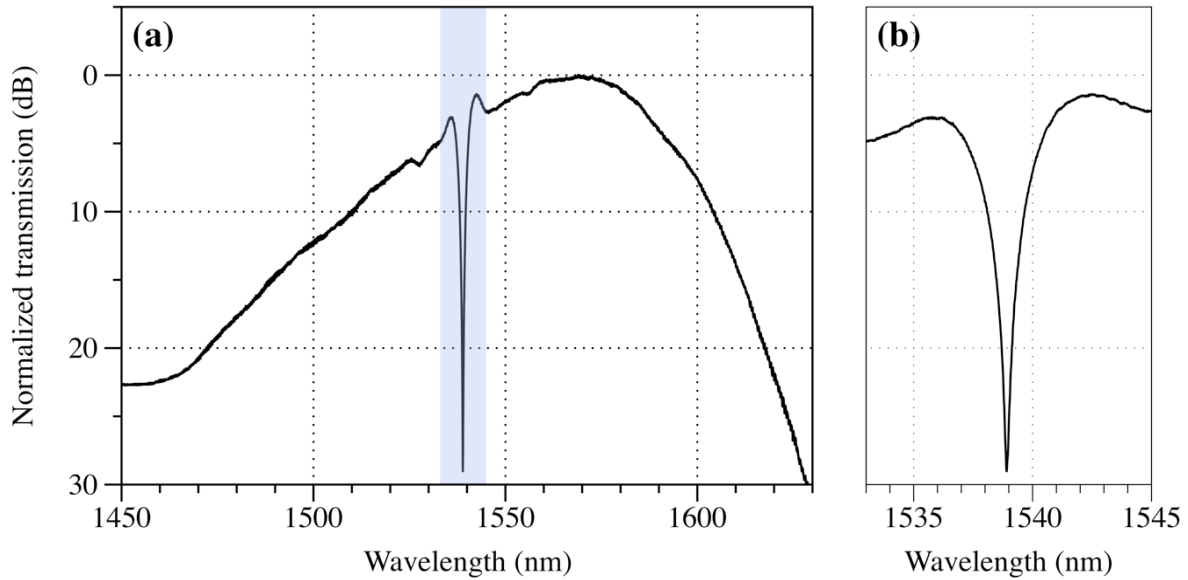

**Figure 5.** Transmission as a function of the wavelength (a). The resonant dip is clearly visible at a wavelength of 1539 nm. Magnified resonance behavior of the ridge resonator (b).

**Figure 6(a)** shows the transmission of the ridge resonator as a function of wavelength for substrate temperatures of 30, 50, 70 and 90°C. It can be seen that the resonance wavelength is remarkably stable over this temperature range, exhibiting a slight shift to longer wavelengths when the substrate temperature is increased, but with the resonance remaining within its 3dB bandwidth over the entire temperature range. **Figure 6(b)** presents the measured resonant wavelengths for substrate temperatures ranging between 30 and 90°C. The fitted linear function indicates that the resonance frequency shifts by around 19.5 pm/°C, which matches well with the simulated temperature sensitivity of ~21.0 pm/°C (see **Supporting Information**, **Section 4**).

Our theoretical analysis indicates that the resonant wavelength of the ridge resonator depends on the ratio of the TM bound mode and TE slab mode effective indices. In the current configuration, the TM mode effective index increases with temperature slightly quicker than the TE slab mode effective index. The temperature insensitivity could be further improved by engineering the TM ridge mode propagation characteristics. The TM mode of the ridge resonator has a strong evanescent field outside of the waveguide as it can be seen in **Figure 2(c)**. This is in contrast to the TE mode which is tightly bound. Hence, the evanescent field can be used to engineer the properties of the TM ridge mode relative to the TE slab. The ridge could be coated with material with a negative thermo-optic coefficient.[29,30] For example, when the ridge is coved with 105 nm of $TiO_2$ with a thermo-optic coefficient of $-10^{-4}$ $K^{-1}$, athermal ridge resonance can be achieved. The resulting temperature insensitivity could make these ridge resonators highly attractive for many practical applications in integrated silicon photonics.

Another attractive possibility is to use the strong evanescent field of the ridge resonance for sensing applications such as biophotonics,[31] as a refractive index change at the surface would result in a large change in the properties of the TM mode, but minimal change in the TE slab resulting in a shift of the resonance wavelength. Due to the single resonance over a broad wavelength range, this shift can be unambiguously detected unlike other types of resonators that exhibits cyclical resonances, resulting in increasing sensing dynamic range.

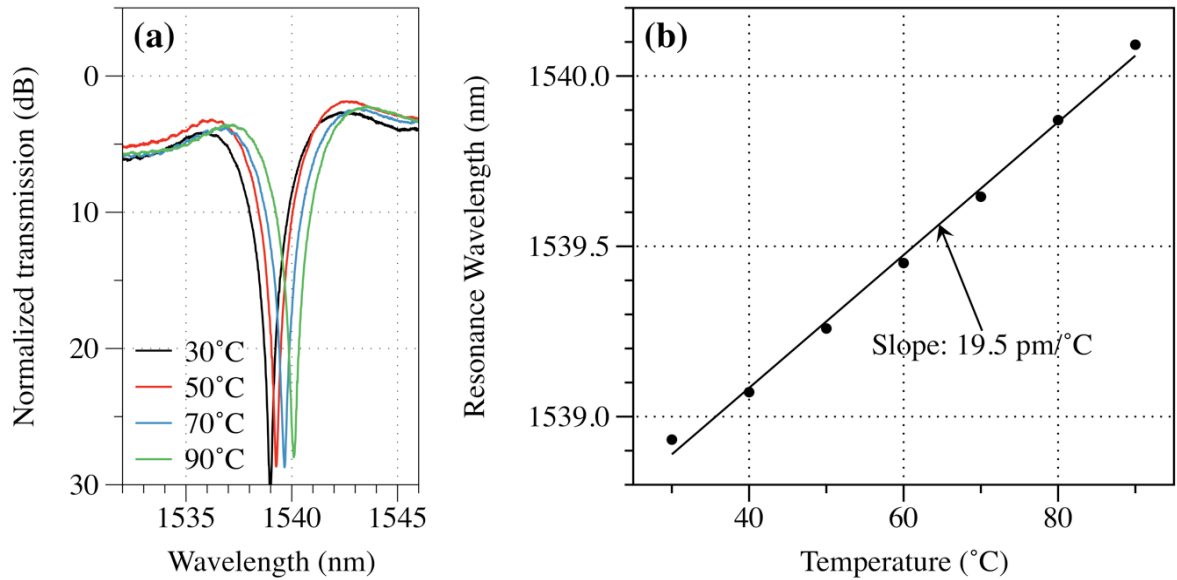

**Figure 6.** Transmission as a function of wavelength for a ridge resonator for different temperatures, highlighting the temperature insensitive behavior of the ridge resonator (a). The measured resonance wavelengths as a function of substrate temperature is fitted with a linear function to determine a temperature sensitivity of 19.5 pm/°C (b).

In our demonstration, the linewidth of the resonance and the extinction ratio is limited by the parabolic mirror structures used to generate and collect the TE beams in the silicon slab. Our rigorous simulation and analytical analysis in Sections 3 and 4 show that it should be possible to achieve high extinction ratio as well as very narrow linewidth resonances when the ridge width is close to 'magic' width. In order to realize such a narrow linewidth resonance, the ridge must be relatively long and must be excited with a very broad TE beam. Methods to practically achieve such high quality and broad beam excitation are currently under investigation.

## 6. Conclusion

We have conceived and experimentally demonstrated means by which bound state in the continuum phenomenon can be used to achieve strong resonance in a silicon photonic waveguide platform. We have presented theory which describes this effect as a type of Fano resonance and experimentally demonstrated resonators with an extinction ratio of 25 dB and the Q-factor of 520 and are highly temperature stable. This proof of concept demonstration is un-optimized, and we believe these structures can achieve higher Q and even more temperature stable operation with further optimization. We believe that this is the first experimental demonstration of such behavior in any waveguide platform, but is of particular significance as our platform satisfies silicon photonic foundry design rules. We anticipate that this type of resonator could form the basis of a new class of wavelength dependent silicon photonic components with particular relevance for integrated transceivers for data communications applications based on coarse wavelength multiplexing. We also anticipate that this new ridge resonator could be a valuable component for photonic chip sensors and high extinction filtering applications such as required for integrated quantum optics where external thin-film filters are currently indispensable.

**Supporting Information**

Additional supporting information may be found in the online version of this article at the publisher's website.

**Acknowledgements**

The authors acknowledge the generous support of the ARC Centre of Excellence CUDOS (CE110001018) and ARC discovery project (DP150101336). Furthermore, the authors acknowledge the facilities, and the scientific and technical assistance, of the Micro Nano Research Facility (MNRF) and the Australian Microscopy & Microanalysis Research Facility at RMIT University. This work was performed in part at the Melbourne Centre for Nanofabrication (MCN) in the Victorian Node of the Australian National Fabrication Facility (ANFF).

**References**

[1]     W. Bogaerts, S. K. Selvaraja, P. Dumon, J. Brouckaert, K. De Vos, D. Van Thourhout, and R. Baets, IEEE J. Select. Topics Quantum Electron. **16**, 33 (2010).

[2]     Q. Xu, B. Schmidt, S. Pradhan, and M. Lipson, Nature, **435**, 7040 (2005).

[3]     T. Komljenovic, S. Srinivasan, E. Norberg, M. Davenport, G. Fish, and J. E. Bowers, IEEE J. Select. Topics Quantum Electron. **21**, 214 (2015).

[4]     D. C. Marinica, A. G. Borisov, and S. V. Shabanov, Phys. Rev. Lett. **100**, 465 (2008).

[5]     C. W. Hsu, B. Zhen, A. D. Stone, J. D. Joannopoulos, and M. Soljačić, Nat. Rev. Mater. **1**, 16048 (2016).

[6]     K. Koshelev, S. Lepeshov, M. Liu, A. Bogdanov, and Y. Kivshar, Phys. Rev. Lett., **121**, 19, 193903 (2018).

[7]     M. Liu and D.-Y. Choi, Nano Lett., **18**, 12, 8062–8069 (2018).

[8]     K. Koshelev, A. Bogdanov, and Y. Kivshar, Science Bulletin, 1–11 (2018).

[9]     S. Romano, G. Zito, S. Torino, G. Calafiore, E. Penzo, G. Coppola, S. Cabrini, I. Rendina, and V. Mocella, Photon. Res., **6**, 7, 726–8 (2018).

[10]    Y. Plotnik, O. Peleg, F. Dreisow, M. Heinrich, S. Nolte, A. Szameit, and M. Segev, Phys. Rev. Lett. **107**, 183901 (2011).

[11]    Y. Boretz, G. Ordonez, S. Tanaka, and T. Petrosky, Phys. Rev. A **90**, 023853 (2014).

[12]    C. W. Hsu, B. Zhen, J. Lee, S. Chua, S. G. Johnson, J. D. Joannopoulos, and M. Soljačić, Nature **499**, 188–191 (2013).

[13]    C. Zou, J. Cui , F. Sun, X. Xiong, X. Zou , Z. Han, and G. Guo, Laser & Photonics Review **9**, 114-119 (2015).

[14] R. Pafchek, R. Tummidi, J. Li, M. A. Webster, E. Chen, and T. L. Koch, Appl. Opt. **48**, 958 (2009).

[15] G. Li, J. Yao, Y. Luo, H. Thacker, A. Mekis, X. Zheng, I. Shubin, J.-H. Lee, K. Raj, J. E. Cunningham, and A. V. Krishnamoorthy, Opt. Exp. **20**, 12035 (2012).

[16] M. A. Webster, R. M. Pafchek, A. Mitchell, and T. L. Koch, IEEE Photon. Technol. Lett. **19**, 429 (2007).

[17] T. Ako, A. Hope, T. Nguyen, A. Mitchell, W. Bogaerts, K. Neyts, and J. Beeckman, Opt. Exp. **23**, 2846 (2015).

[18] A. P. Hope, T. G. Nguyen, A. Mitchell, and W. Bogaerts, IEEE Photon. Technol. Lett. **28**, 493 (2016).

[19] T. G. Nguyen, R. S. Tummidi, T. L. Koch, and A. Mitchell, Opt. Lett. **34**, 980-982 (2009).

[20] T. G. Nguyen, R. S. Tummidi, T. L. Koch, and A. Mitchell, Opt. Express **18**, 7243-7252 (2010).

[21] R. S. Tummidi, T. G. Nguyen, and A. M., Group IV Photonics 104 (2011).

[22] T. G. Nguyen, R. S. Tummidi, T. L. Koch, and A. Mitchell, IEEE Photon. Technol. Lett. **21**, 486 (2009).

[23] E. A. Bezus, D. A. Bykov, and L. L. Doskolovich, Photon. Res. **6**, 1084-1093 (2018).

[24] K. X. Wang, Z. Yu, S. Sandhu, and S. Fan, Opt. Lett. **38**, 100 (2013).

[25] S. Fan, W. Suh, and J. D. Joannopoulos, J. Opt. Soc. Am. **A20**, 569-572 (2003).

[26] P. Dumon, W. Bogaerts, R. Baets, J. M. Fedeli, and L. Fulbert, Electron. Lett. **45**, 12 (2009).

[27] A. P. Hope, T. G. Nguyen, A. Mitchell and W. Bogaerts, IEEE Phot. Tech. Lett. **28**, 4 (2016).

[28] D. Taillaert, W. Bogaerts, P. Bienstman, T. Kraus, P. Van Daele, I. Moeman, S. Verstuyft, K. De Mesel, and R. Baets, IEEE Journal of Quantum Electronics, **38**, 7 (2002).

[29] J. Teng, P. Dumon, W. Bogaerts, H. Zhang, X. Jian, X. Han, M. Zhao, G. Morthier, and R. Baets, Opt. Exp. **17**, 14627 (2009).

[30] B. Guha, J. Cardenas, and M. Lipson, Opt. Exp. **21**, 26557 (2013).

[31] M. C. Estevez, M. Alvarez, and L. M. Lechuga, Laser & Photon. Rev. **6**, 463 (2011).

# Supporting Information

**Ridge Resonance: A new Resonance Phenomenon for Silicon Photonics**

*Thach G. Nguyen[1,*], Guanghui Ren[1], Steffen Schoenhardt[1], Markus Knoerzer[1], Andreas Boes[1], and Arnan Mitchell[1,*]*

*Corresponding Author: E-mail: thach.nguyen@rmit.edu.au, arnan.mitchell@rmit.edu.au

[1]School of Engineering, RMIT University, Melbourne, VIC 3001, Australia

**Abstract:** In this Supporting Information document more details are provided on (1) the theory of the ridge resonance, (2) the rigorous simulation of the ridge resonance, (3) the influence of the grating couplers on the measured spectra and (4) the temperature dependence of the ridge resonance.

1. Theory of Ridge Resonance

In this section, the theory of the resonance in ridge waveguides is developed based on coupled mode theory formalism. Drawing on the analogy to a two-port resonator,[1] the ridge resonance can be considered as the sum of resonance-assisted component due to coupling between TM guided mode in the ridge and the TE slab mode in the slab, and a background component due to Fresnel reflections at the ridge walls as illustrated in **Figure S1**.

When excited, the TM guided mode in the ridge waveguide has the z-dependent amplitude given by $u(z) = u_0 e^{j\frac{2\pi}{\lambda_0}n_{tm}z - \frac{2\pi}{\lambda_0}n'_{tm}z} = e^{j\omega_0 z - \gamma z}$, where $\lambda_0$ is the wavelength, $n_{tm}$ and $n'_{tm}$ are the real and imaginary parts of the complex effective index of the TM guided mode. Here we define the resonant 'frequency' $\omega_0 = \frac{2\pi}{\lambda_0}n_{tm}$ and decay rate $\gamma = \frac{2\pi}{\lambda_0}n'_{tm}$ which is proportional to the leakage loss.

The guided TM mode is excited by incoming TE waves from the slab regions on both sides of the ridge

$$S_+ = \begin{bmatrix} s_{1+} \\ s_{2+} \end{bmatrix}$$

with coupling coefficients

$$K = \begin{bmatrix} \kappa_1 \\ \kappa_2 \end{bmatrix}$$

When excited, the TM guided mode leaks to outgoing TE waves

$$S_- = \begin{bmatrix} s_{1-} \\ s_{2-} \end{bmatrix}$$

on both sides of the waveguides with the coupling coefficients

$$D = \begin{bmatrix} d_1 \\ d_2 \end{bmatrix}$$

The dynamic equations for the amplitude $u$ of the TM guided mode can be written as

$$\frac{du}{dz} = (j\omega_0 - \gamma)u + (K^T)^* S_+ \tag{S1}$$

$$S_- = CS_+ + uD \tag{S2}$$

where $C$ is the scattering matrix of the non-resonant assisted coupling between incoming and outgoing TE waves. This scattering matrix has the following form

$$C = e^{j\phi} \begin{bmatrix} r & jt \\ jt & r \end{bmatrix} \qquad (S3)$$

where $\phi$ is a phase factor, and r and t are the reflection and transmission of the non-resonant assisted coupling process.

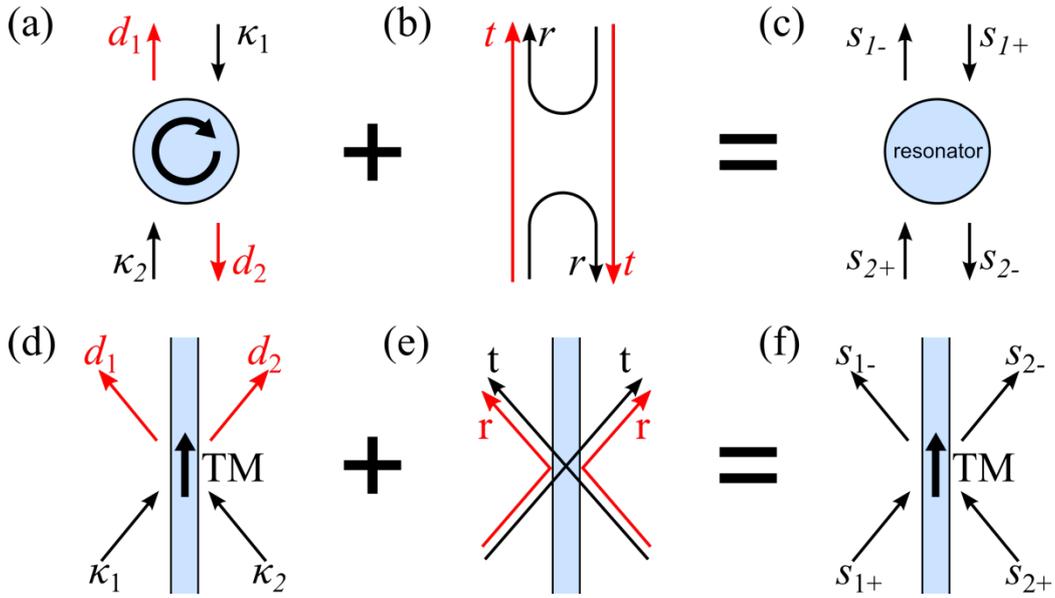

**Figure S1.** A two-port resonator (c), can be described as the sum of resonance-assisted component (a) and a background component (b). In the case of a ridge resonator one can apply a similar description: a ridge resonator (f) is the sum of resonance-assisted component (d) and a background component (e), where TM is the transverse magnetic mode excited in the ridge. Figures (a) to (c) are adapted from Ref. [1].

For externally incident TE wave of amplitude $s_{i+}$ at the wavelength $\lambda$ incident on the ridge with incident angle $\theta$, the complex amplitude of the TE wave along the waveguide axis z is defined as $s_{i+}(z) = s_{i0} e^{j\frac{2\pi}{\lambda} n_{te} \sin\theta z} = s_{i0} e^{j\omega z}$, where $n_{te}$ is the effective index of the TE wave. Here we define the excitation 'frequency' $\omega = \frac{2\pi}{\lambda} n_{te} \sin\theta$.

By drawing an analogy to the temporal coupled mode theory in which the variable $z$ is equivalent to time variable, we can obtain the scattering matrix $S$ describing the relationship between incoming TE wave $S_+$ and outgoing TE wave $S_-$ as

$$S_- = SS_+ = \left[C + \frac{D(K^T)^*}{j(\omega-\omega_0)+\gamma}\right]S_+ \quad \text{(S4)}$$

Using the energy-conservation and reciprocal rules and following the same procedures as in,[1,2] the relationships between coupling matrices *K*, *D* and *C* can be derived as the followings:

$$K = D \quad \text{(S5)}$$

$$CD^* = -D \quad \text{(S6)}$$

$$D^T D = 2\gamma \quad \text{(S7)}$$

Because of the symmetric of the waveguide structure, we further have $d_1 = d_2 = \sqrt{\gamma}$.

Using these relationships, the overall system scattering matrix is then given by

$$S = e^{j\phi}\left\{\begin{bmatrix} r & jt \\ jt & r \end{bmatrix} + \frac{\gamma}{j(\omega-\omega_0)+\gamma}\begin{bmatrix} -(r+jt) & -(r+jt) \\ -(r+jt) & -(r+jt) \end{bmatrix}\right\} \quad \text{(S8)}$$

The reflection coefficient *R* of an incident TE wave is therefore

$$R \equiv S_{11} = e^{j\phi}\frac{j[(\omega-\omega_0)r - \gamma t]}{j(\omega-\omega_0)+\gamma} \quad \text{(S9)}$$

When ignoring the non-resonant assisted coupling process, i.e. *r* = 0 and *t* = 1, the reflection coefficient only due to resonant-assisted process can be written as

$$R = \frac{-j\gamma}{j(\omega-\omega_0)+\gamma} \quad \text{(S10)}$$

which is completely determined by the leakage loss of the guided TM mode, the effective indices of the guided TM mode and TE slab mode in the slab region.

To validate the theory, we compare the theoretical prediction of the reflection coefficient using **Equation (S9)** to the rigorous simulation results of the reflection coefficient using Mode Matching simulation as described in **Section S2**. The effective index and leakage loss of the guide TM mode were found from the eigensolution using the Mode-matching eigen mode solver.[3] To calculate the background reflection coefficient due to non-resonant assisted process, we first calculated the reflection and transmission coefficients at the cladding/ridge interface of the TE waves incident from the cladding and ridge regions use the same method as described in Ref [3]. Then the non-resonant assisted reflection coefficient was calculated as:

$$r = \left| r_{12} + \frac{r_{21} t_{12} t_{21} e^{j2\delta}}{1 - r_{21}^2 e^{j2\delta}} \right| \quad \text{(S11)}$$

Where $r_{12}$ and $t_{12}$ are the reflection and transmission coefficients of the TE wave at the cladding/ridge interface incident from the cladding region, $r_{21}$ and $t_{21}$ are the reflection and transmission coefficients at the ridge/cladding interface of the TE wave incident from the ridge region, $\delta = \frac{2\pi}{\lambda_0} n_{te}^{core} w \cos\varphi$, $n_{te}^{core}$ is the effective index of TE wave inside the ridge, $\varphi$ is the propagating angle of the TE wave inside the ridge and $w$ is the ridge width.

**Figure S2** and **Figure S3** shows the transmission and reflection coefficients of an incident TE wave as a function of wavelength calculated with rigorous simulation and analytical theory, for a resonant wavelength of 1.55 μm (**Figure S2**) and 1.31 μm (**Figure S3**), respectively. Also shown in **Figure S2** and **Figure S3** are the background transmission and reflection of the non-resonant assisted process and the transmission/reflection coefficient due to the resonance alone using **Equation (S10)**. It can be seen that the theoretical prediction when including the resonant-assisted reflection and background reflection (analytical case) matches very well to the rigorous simulation results in both cases (strong and weak background). It can be seen from **Figure S3** that when the background non-resonant assisted reflection is strong, the background reflection can interfere with the reflection only due to resonant-assisted process resulting in Fano resonance.[4]

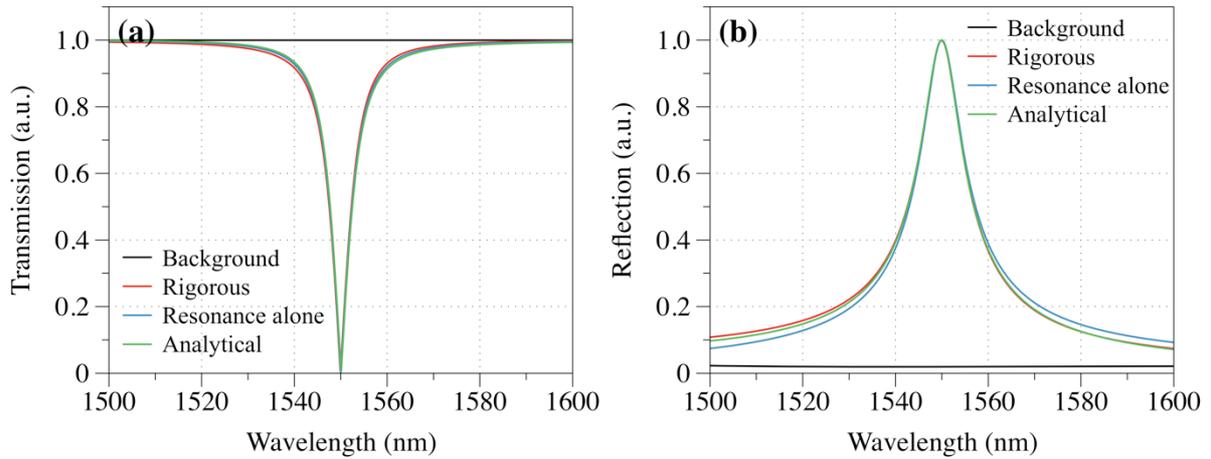

**Figure S2.** Ridge resonator with a designed resonance at 1.55 μm. (a) and (b) present the transmission and reflection spectra for the Fresnel background, rigorous simulation, only due to resonant-assisted process and the analytical solution which is the combination of the resonant-assisted response with the background.

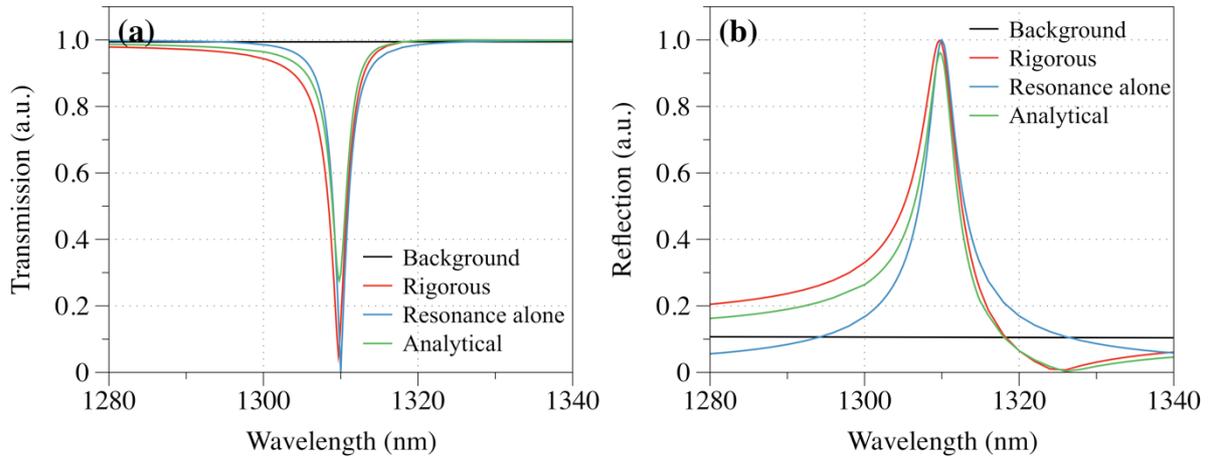

**Figure S3.** Ridge resonator with a designed resonance at 1.31 μm. (a) and (b) present the transmission and reflection spectra for the Fresnel background, rigorous simulation, only due to resonant-assisted process and the analytical solution which is the combination of the resonant-assisted response with the background.

## 2. Rigorous Simulation of Ridge Resonance

A fully vectorial eigenmode expansion [5] technique was employed to calculate the response of the ridge to an incident TE beam from one side of the ridge. The structure was divided into a number of uniform multilayer slab sections. In each slab section, the total electromagnetic field was expressed as a superposition of both forward and backward propagating slab modes, including guided and radiations modes, for both TE and TM polarizations of the corresponding slab. These slab modes propagate at different angles, but they must be phase-matched along the ridge axis. The continuum of the slab radiation modes was discretized by placing the ridge structure vertically between two perfectly conducing planes. By matching the fields on each slab region at the ridge boundaries, the scattering matrices of the ridge boundaries can be found. Cascading the scattering matrices of all ridge boundaries and the slab regions gives the scattering matrix of the whole structure. From the scattering matrix, the reflection and transmission coefficients of the ridge when excited with an oblique TE wave can be readily found.

3. Spectral Influence of Grating Coupler

**Figure S4** shows the transmission as a function of wavelength for the slab mode launching structure using parabolic reflectors without a resonator and a standard silicon wire waveguides for comparison. One can see that the shape of the spectra for the wire waveguide is similar to the one when a TE slab mode is launched, indicated that the overall spectral behavior is dominated by the wavelength dependent coupling efficiency of the grating couplers. One can further see that using the parabolic reflector to launch and collect the TE slab mode introduces an additional loss of approximately 6 dB.

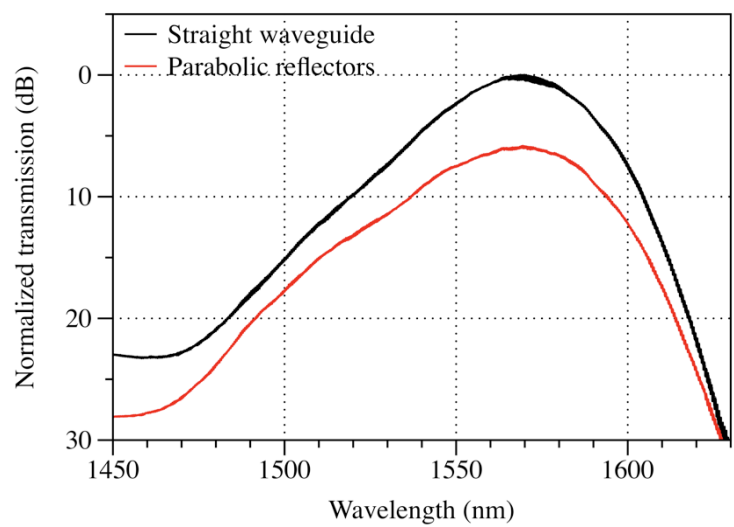

**Figure S4.** Transmission as a function of wavelength for the slab mode launching structure without a ridge resonator and a standard silicon wire waveguide for comparison.

4. Temperature Dependence of Resonance

At the resonant wavelength, the TE mode - incident under an angle in the etched slab - and the TM mode in the unetched ridge are phase matched. As the TE and TM modes show different dispersion, this can result in a thermal shift of resonant wavelength.

An eigenmode solver based on mode-matching method [3] was used to predict the thermal behavior of the ridge resonator. The incident angle of the TE mode, which is dictated by the phase matching condition, was calculated for a nominal resonant wavelength (i.e. wavelength at 30°C). While keeping the incident angle constant, the material refractive indices of silicon and silica in the mode solver model were changed according to their respective thermo-optic coefficients,[6,7] to simulate changes in temperature. By recording the dispersion of the TE mode incident under a fixed angle and the TM mode for different temperatures, the thermal shift of the resonance can be found at the crossings for the dispersion curves for different temperatures (as indicated in **Figure S5**). The temperature sensitivity of the resonant wavelength was found to be ~21 pm/°C.

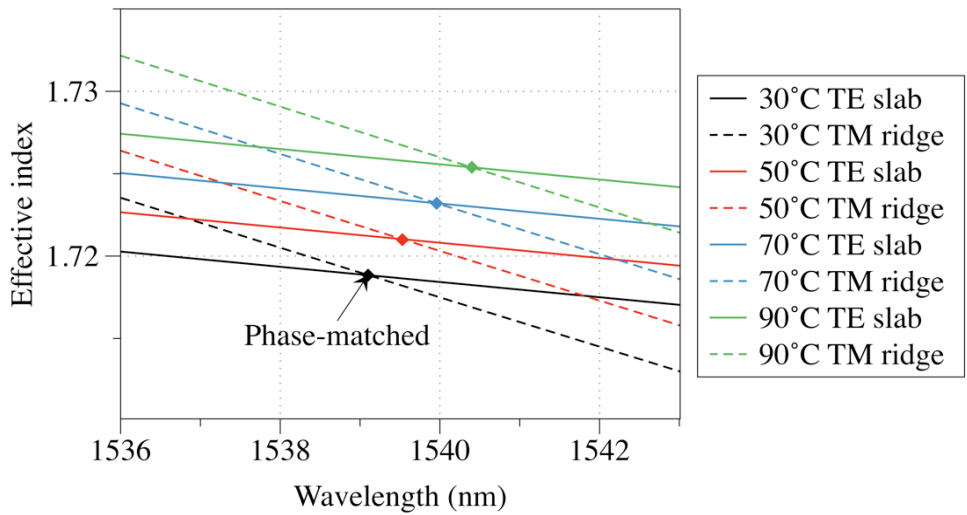

**Figure S5.** TE and TM mode dispersion for different temperatures for a ridge resonator with resonant wavelength at 1539.1 nm at 30°C.

We also investigated how the temperature sensitivity of the resonance wavelength depends on the resonance wavelength. The temperature shift of the resonance wavelength was simulated for different wavelengths from 1500 to 1590 nm and is presented in **Figure S6**. One can see that temperature sensitivity of the ridge resonance is lower for longer resonance wavelengths.

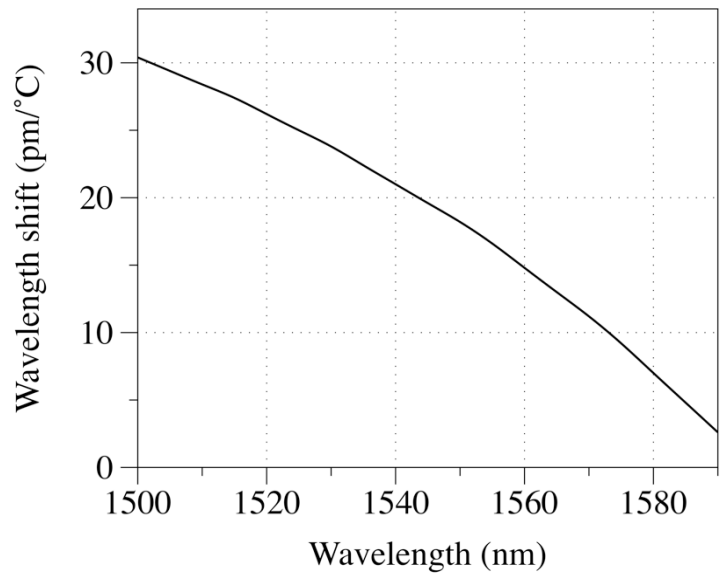

**Figure S6.** Simulated temperature shift of the resonance wavelength as a function of the wavelength at a temperature of 25°C.

**References**

[1]  K. X. Wang, Z. Yu, S. Sandhu, and S. Fan, Opt. Lett. **38**, 100-102 (2013).

[2]  S. Fan, W. Suh, and J. D. Joannopoulos, J. Opt. Soc. Am. **A20**, 569-572 (2003).

[3]  T. G. Nguyen, R. S. Tummidi, T. L. Koch, and A. Mitchell, IEEE Photon. Technol. Lett. **21**, 486 (2009).

[4]  K. X. Wang, Z. Yu, S. Sandhu, and S. Fan, Opt Lett **38**, 100 (2013).

[5]  P. Bienstman, PhD Thesis, Ghent University (2001).

[6]  G. Cocorullo, F. G. Della Corte, and I. Rendina, Appl. Phys. Lett. **74**, 3338 (1999).

[7]  J. M. Jewell, J. Am. Ceram. Soc. **74**, 1689 (1991).